\documentstyle[aps,twocolumn,prl,epsf,amssymb]{revtex}

\def\R{{\Bbb{R}}}
\def\Z{{\Bbb{Z}}}
\def\T{{\Bbb{T}}}
\def\H{\hat{\cal H}}
\def\U{\hat{\cal U}}
\def\px{\hat{p}_x}
\def\py{\hat{p}_y}

\newcommand{\bra}[1]{\langle #1|}
\newcommand{\ket}[1]{|#1\rangle}
\newcommand{\braket}[2]{\langle #1|#2\rangle}
\newcommand{\half}{{\textstyle{\frac{1}{2}}}}
\newcommand{\ihalf}{{\textstyle{\frac{i}{2}}}}
\newcommand{\shalf}{{\scriptstyle{\frac{1}{2}}}}

\newcommand{\quar}{{\textstyle{\frac{1}{4}}}}
\newcommand{\eigh}{{\textstyle{\frac{1}{8}}}}
\newcommand{\oneL}{{\textstyle{\frac{1}{L}}}}
\newcommand{\ad}{{\rm ad\,}}
\newcommand{\tr}{{\rm tr\,}}
\begin{document}
\draft
\twocolumn[\hsize\textwidth\columnwidth\hsize\csname %
@twocolumnfalse\endcsname
\title{A map from 1d Quantum Field Theory to Quantum Chaos on a 2d Torus}
\author{Toma\v z Prosen}
\address{Physics Department, Faculty of Mathematics and Physics,
University of Ljubljana, Jadranska 19, 1111 Ljubljana, Slovenia
}  
\date{\today}
\draft
\maketitle
\begin{abstract}
\widetext
Dynamics of a class of quantum field models on 1d lattice in Heisenberg
picture is mapped into a class of `quantum chaotic' one-body systems on
configurational 2d torus (or 2d lattice) in Schr\" odinger picture.
Continuum field limit of the former corresponds to quasi-classical
limit of the latter.
\end{abstract}

\pacs{PACS numbers: 05.45.+b, 05.30.Fk, 03.65.Fd, 72.10.Bg}
]
\narrowtext
Understanding of long-time dynamics of interacting quantum many-body
systems or quantized fields is a long standing open problem.
In particular, one would like to understand the conditions for the
emergence of quantum mixing (implying ergodicity), i.e. general decay of time 
autocorrelation functions. Inspired by rich quantum behaviour 
of non-integrable few body systems having (partially) chaotic classical 
limit\cite{QChaos} few papers appeared recently concerning `quantum chaos' 
in non-integrable many-body systems \cite{JLP,MBQC,KtV}. In \cite{KtV}
{\em dynamical phase transition} from non-ergodic dynamics exhibiting ideal 
transport to mixing dynamics exhibiting normal transport has been demonstrated
numerically in a non-integrable kicked t-V fermion model on 1d lattice.

Below we construct an exact {\em linear mapping} from dynamics
of a certain large class of interacting infinite spin$-\half$ chains in
Heisenberg picture to a class of non-linear {\em One-Body Image dynamical
systems} (OBI) in Schr\" odinger picture which are realized
either on configurational 2d torus or on 2d lattice (tight-binding (TB)
formulation). Further we will show by working out two examples: (i) how
integrable behavior of the infinite {\em XX spin chain in spatially modulated
transversal magnetic field} is connected to the integrability of OBI and
to the Harper equation \cite{Harper}, and (ii) demonstrate the phase
transition of the related {\em non-integrable kicked XX chain} from
non-ergodic dynamics to ergodic and mixing dynamics corresponding to the
stochastic transition from regular to chaotic motion of OBI in the classical
limit.

Let $\sigma^s_j,j\in\Z,s\in\{x,y,z\}$ denote a chain of independent Pauli
spin$-\half$ variables. We start by generalizing the result of \cite{KI}, 
namely we find that the operator space spanned by the following
set of spatially modulated observables 
$\{U_n(\vartheta),V_n(\vartheta); n\in\Z,\vartheta\in[-\pi,\pi)\}$
\begin{eqnarray*}
U_n(\vartheta) &=&\sum_{j=-\infty}^{\infty}e^{i(j+\shalf|n|)\vartheta}
\cases{
\sigma^x_j\,(\sigma^z_j)_{n-1}\,\sigma^x_{j+n} & $n\ge 1$,\cr
-\sigma^z_j & $n = 0$,\cr
\sigma^y_j\,(\sigma^z_j)_{-n-1}\,\sigma^y_{j-n} & $n\le-1$,\cr}\\
V_n(\vartheta) &=&\sum_{j=-\infty}^{\infty}e^{i(j+\shalf|n|)\vartheta}
\cases{
\sigma^x_j\,(\sigma^z_j)_{n-1}\,\sigma^y_{j+n} & $n\ge 1$, \cr
 1 & $n = 0$,\cr
-\sigma^y_j\,(\sigma^z_j)_{-n-1}\,\sigma^x_{j-n} & $n\le-1$. \cr}
\end{eqnarray*}
where $(\sigma^z_j)_k := \prod_{l=1}^k \sigma^z_{j+l}$ for $k\ge 1$ and
$(\sigma^z_j)_0:=1$,
is closed under the Lie bracket $[A,B]=AB-BA$ and
forms an infinitely dimensional Dynamical Lie Algebra (DLA)
\begin{eqnarray}
\lbrack U_{n}(\vartheta),U_{l}(\varphi) \rbrack
&=& 2i \exp\left(\ihalf(l\vartheta+n\varphi)s_{l-n}\right)V_{n-l}(\vartheta+\varphi) 
\nonumber \\
&-& 2i \exp\left(\ihalf(l\vartheta+n\varphi)s_{n-l}\right)V_{l-n}(\vartheta+\varphi),
\nonumber \\
\lbrack U_{n}(\vartheta),V_{l}(\varphi) \rbrack
&=& 2i \exp\left(\ihalf(-l\vartheta+n\varphi)s_l\right)U_{n+l}(\vartheta+\varphi) 
\nonumber\\
&-& 2i \exp\left(\ihalf(l\vartheta+n\varphi)s_l\right)U_{n-l}(\vartheta+\varphi),
\nonumber\\
\lbrack V_n(\vartheta),V_l(\varphi) \rbrack &=& (s_n+s_l)\Bigl\{
\sin\left(\half(l\vartheta+n\varphi)\right)\times \nonumber \\
\bigl(
(s_{n-l}s_l+1)&V_{n-l}&(\vartheta+\varphi) - (s_{l-n}s_l+1)V_{l-n}(\vartheta+\varphi)
\bigr) \nonumber \\
+\;\; 2\sin\bigl(\half(l\vartheta\!\!&-&\!\!n\varphi)\bigr)V_{n+l}(\vartheta+\varphi)
\Bigr\}. \label{eq:DLA}
\end{eqnarray}
where $s_n:=-1,0,1$ for $n<,=,>0$, resp., is a sign of integer $n$.
Few notable members of DLA are: Ising or XX hamiltonian
$H_I=J U_1(0)$,$H_{XX} = J (U_1(0)+U_{-1}(0))$,
spin interaction with modulated transversal magnetic field 
$h_z=h\cos(\epsilon j)$ with period $2\pi/\epsilon$ lattice spacings
$H_{mh} = \half h (U_0(\epsilon) + U_0(-\epsilon))$, 
spin current $j_s = V_{1}(0) + V_{-1}(0)$, etc.
Let us fix the {\em fundamental field modulation} $\epsilon$  
and introduce the following notation:
\begin{eqnarray*}
(n,k)\in\Z^2,\qquad\quad
U^\pm_{n,k} &=& \half\left( U_n(k\epsilon) \pm U_{-n}(-k\epsilon)\right),\\
V^+_{n,k} &=& \half\left( V_n(k\epsilon) + V_{-n}(-k\epsilon)\right)s_n,\\
V^-_{n,k} &=& \half\left( V_{n}(k\epsilon) - V_{-n}(-k\epsilon)\right),\\
W^\pm_{n,k} &=& U^\pm_{n,k} + i V^\pm_{n,k},\\
(y,x)\in\T^2,\quad
W^\pm(y,x) &=& \frac{1}{2\pi}\sum_{n,k=-\infty}^\infty 
e^{i(ny+kx)} W^\pm_{n,k}.
\end{eqnarray*}
We may also consider $\epsilon$ as a {\em lattice spacing} and treat
$W^\pm_{n,k}$ as a set of spatially $2\pi$-periodic fields.
We will assume that the modulation is {\em incommensurable} with the
lattice spacing, i.e. that $\epsilon/2\pi$ is {\em irrational},
otherwise obsrvables $W^\pm_{n,k}$ are periodic w.r.t. index $k$.
DLA becomes a Hilbert space when we introduce an infinite temperature 
(grand) canonical scalar product \cite{KI} $(A|B) := 
\lim_{L\rightarrow\infty}\oneL 2^{-L} \tr A^\dagger B$,
where $L$ is a diverging length of the spin-chain.
Let the two linear subspaces spanned by $W^\sigma_{n,k}$ 
(or $W^\sigma(y,x)$) for $\sigma\in\{+,-\}$
be denoted by ${\frak M}_\sigma$.
The spaces ${\frak M}_+$ and ${\frak M}_-$ are {\em orthogonal}
and observables $W^+_{n,k}$ ($n,k\in\Z$) and $W^-_{n,k}$ (for $n\ge 1$) 
form orthonormal bases in each of them, since one can show 
\begin{eqnarray}
(W^+_{n,k}|W^+_{m,l}) &=& \delta_{n,m} \delta_{k,l},\quad
(W^+_{n,k}|W^-_{m,l}) = 0,\nonumber\\
(W^-_{n,k}|W^-_{m,l}) &=& (\delta_{n,m}-\delta_{n,-m})\delta_{k,l}.
\label{eq:scal}
\end{eqnarray}
The full set $\{W^-_{n,k}\}$ is over-complete, since $W^-_{-n,k}=-W^-_{n,k}$,
while the subspace ${\frak M}_+={\frak M}^\dagger_+$ is self-adjoint,
since $W^{+\dagger}_{n,k}=W^+_{-n,-k}$. One can write analogous relations
in terms of continuous variables $(y,x)$.
We have $\text{DLA} =
{\frak M}^\dagger_- \oplus {\frak M}_+ \oplus {\frak M}_-$.
Note that the {\em adjoint map} $(\ad A)B=[A,B]$ generates the
Heisenberg motion on DLA, 
$\exp(it\ad A)B=e^{i t A} B e^{-i t A}$. In particular, the motion generated
by $U^+_{n,k}$ has a beautiful structure. Let us write the 
self-adjoint Hamiltonian in a general form as
\begin{equation}
H = \sum_{n,k} \quar g_{n,k} \left(
U^{+}_{n,k} e^{i\gamma_{n,k}} + U^{+}_{n,-k} e^{-i\gamma_{n,-k}}
\right)
\label{eq:H}
\end{equation}
using two sets of possibly time dependent real coefficients 
$g_{n,k}=g_{n,k}(t),\gamma_{n,k}=\gamma_{n,k}(t)$.
Tedious but straightforward calculation, using algebra (\ref{eq:DLA}), gives
the action of $\ad H$ on two continuous sets of
observables $W^+(y,x),W^-(y,x)$, $(y,x)\in \T^2$ which can be written in
terms of two non-local `Schr\" odinger operators' $\H^\pm$
\begin{eqnarray}
(\ad H)&&W^\pm(y,x) = 
- {\textstyle\frac{1}{\hbar}}\H^\pm W^\pm(y,x), \label{eq:adH} \\
\H^+ = \sum_{n,k}\hbar g_{n,k}&\bigl(&
\sin(n\px\!-\!k\py)\sin(kx\!+\!ny\!-\!\gamma_{n,k}) 
\nonumber\\ \noalign{\vskip -0.18in}
-&& \sin(n\px\!+\!k\py)\sin(kx\!-\!ny\!-\!\gamma_{n,k})\bigr),\nonumber\\
\H^- = \sum_{n,k}\hbar g_{n,k}&\bigl(&
\cos(n\px\!-\!k\py)\cos(kx\!+\!ny\!-\!\gamma_{n,k})
\nonumber\\ \noalign{\vskip -0.18in}
+&& \cos(n\px\!+\!k\py)\cos(kx\!-\!ny\!-\!\gamma_{n,k})\bigr),\label{eq:H1}
\end{eqnarray}
where $\hat{p}_{x,y} = -i\hbar\partial/\partial_{x,y}$
are momentum operators conjugate to $x,y$ with an 
`effective Planck constant'\cite{hb}
\begin{equation}
\hbar = \half \epsilon. \label{eq:hbar}
\end{equation}
Since Heisenberg dynamics generated by $H$ is closed on ${\frak M}_\sigma$,
$(\ad H){\frak M}_\sigma \subseteq {\frak M}_\sigma$, one may write a
general time-evolving operator $A(t)\in{\frak M}_\sigma$ in terms of 
a complex-valued `Schr\" odinger wave function', in either `momentum'
$\Psi^A_{n,k}(t)$ or `position' $\Psi^A(y,x;t)$
representation
$$
A(t) = \sum_{n,k} \Psi^A_{n,k}(t)^* W^\sigma_{n,k} = 
\int_{\T^2}\!\!dy dx \Psi^A(y,x;t)^* W^\sigma(y,x).
$$
By means of eq. (\ref{eq:adH}) and the fact that $\H^\sigma$ is Hermitian
on $L^2(\T^2)$ (which can be checked directly using the expressions
(\ref{eq:H1})) one can easily show that the Heisenberg evolution of the
observable $A(t)$, $(d/dt)A(t) = i(\ad H)A$, is {\em fully
equivalent} to the Scr\" odinger equation 
\begin{equation}
i\hbar\frac{d}{dt}\Psi^A(y,x;t) = \H^\sigma \Psi^A(y,x;t).
\label{eq:Sch}
\end{equation}
governing time evolution of one particle on a torus $\T^2$ (OBI).
The {\em bilinear} map $(H,A(t))\leftrightarrow (\H^\sigma,\Psi^A(y,x;t))$ is a
central result
of this Letter. To conclude a general exposition we make few remarks:
(i) A non-trivial `classical limit' $\hbar\rightarrow 0$ of OBI
exists, being {\em equivalent} (\ref{eq:hbar}) 
to the continuum field limit of the quantum spin chain model 
$\epsilon\rightarrow 0$, if $\hbar g_{n,l}$ (and not 
$g_{n,l}$ alone) are kept constant and finite.
(ii) The operators $\H^+$ and $\H^-$ (\ref{eq:H1}) commute
\begin{equation}
[\H^+,\H^-] \equiv 0,
\label{eq:com}
\end{equation}
and the Poisson bracket of the corresponding classical counterparts vanishes.
(iii) As a consequence of the previous remark we find that OBI (\ref{eq:Sch})
(and its classical limit) is {\em integrable}, $\H^{-\sigma}$ being the second 
integral of motion, provided the original spin-field Hamiltonian
$H$ (\ref{eq:H}) or OBI Hamiltonian $\H^\sigma$ is 
autonomous, i.e. $(\partial/\partial t) H\equiv 0$. However, one has
a possibility of chaotic motion in classical limit and emergence
of `quantum chaos' when the problem is explicitly time-dependent,
say that coefficients are periodic functions, $g_{n,k}(t+1)=g_{n,k}(t)$.
In such case one integrates the evolution over one period of time 
and defines the unitary Floquet maps $U=\hat{\cal T}\exp(-i\int_0^1 dt H(t))$,
$\U^\sigma=\hat{\cal T}\exp(-i\int_0^1 dt \H^\sigma(t))$.
(iv) Temporal correlation functions of the
quantum field problem are mapped (using eqs. (\ref{eq:scal})) 
onto transition amplitudes of OBI
\begin{equation}
(A(t)|B(t')) =
\cases{\braket{\Psi^A(t)}{\Psi^B(t')} & $\sigma=+$,\cr
       \braket{\Psi^A(t)}{\hat{\cal P}_y\Psi^B(t')} &
$\sigma=-$,\cr}
\label{eq:CF}
\end{equation}
where $\hat{\cal P}_y\Psi(y,x)=\Psi(y,x)-\Psi(-y,x)$.
Therefore, ergodic properties of many-body dynamics on DLA are determined 
by the spectral properties of OBI: (a) Spin chain is {\em quantum mixing}
in ${\frak M}_\sigma$,
$\lim_{t\rightarrow\infty}(A(t)|B) = 0$, $A,B\in {\frak M}_\sigma$,
iff the spectrum of OBI Hamiltonian $\H^\sigma$ (or of OBI Floquet map 
$\U^\sigma$) does not have (non-trivial) {\em point} component.
(b) Spin chain is {\em quantum ergodic} in ${\frak M}_\sigma$,
$\lim_{T\rightarrow\infty}T^{-1}\int_0^T dt (A(t)|B) = 0$, iff $0$ (or $1$)
is {\em not} in the non-trivial point spectrum of $\H^\sigma$ (or $U^\sigma$). 
In autonomous case, $\partial H/\partial t\equiv 0$,
the Hamiltonian $H$ and the {\em trivial} zero-frequency eigenstate, 
$\H^+\Psi^H=-\hbar \Psi^{[H,H]}=0$,
$\Psi^H_{n,k}=\eigh(g_{n,k}e^{-i\gamma_{n,k}}\!+\!g_{n,-k}e^{i\gamma_{n,-k}}
\!+\!g_{-n,k}e^{-i\gamma_{-n,k}}\!+\!g_{-n,-k}e^{i\gamma_{-n,-k}})$ should be
excluded from ${\frak M}_+$ and $L^2(\T^2)$, 
respectively, i.e. $(A|H)=(B|H)=0$.

We apply the above results to work out two interesting examples.
{\em Example I:} {\em XX spin chain in spatially modulated quasi-periodic
transversal magnetic field} $\vec{h}_j=(0,0,h\cos(\epsilon j))$ (XXmh)
\begin{equation}
H = H_{XX} + H_{mh} = J U^+_{1,0} +
\half h (U^+_{0,1} + U^+_{0,-1}).
\end{equation}
Here the Heisenberg dynamics on DLA is governed by the following commuting
one-body problems
\begin{eqnarray}
\H^+ &=& \alpha \sin\px\sin y - \beta\sin\py\sin x,\label{eq:Hex}\\
\H^- &=& \alpha \cos\px\cos y + \beta\cos\py\cos x,\nonumber
\end{eqnarray}
where $\alpha = 2\epsilon J=4\hbar J,\beta = 2\epsilon h=4\hbar h$.
This models are directly related to the electron motion on 2d
rectangular $a\times b$ lattice in a uniform perpendicular magnetic field
$h'$ within the TB approximation \cite{Harper,Sokoloff}.
In the symmetric gauge $\vec{A} = \half h'(-y,x,0)$ the TB problem
with the band energy ${\cal E}(\vec{K})=\frac{\alpha}{2}\cos(a K_1)
+\frac{\beta}{2}\cos(b K_2)$ reads
\begin{eqnarray}
\H \Psi_{n,k} &=& {\textstyle\frac{\alpha}{2}}\left(
e^{i\shalf\epsilon k}\Psi_{n+1,k} + e^{-i\shalf\epsilon k}\Psi_{n-1,k}\right)
\nonumber\\
&+& {\textstyle\frac{\beta}{2}}\left(
e^{-i\shalf\epsilon n}\Psi_{n,k+1} + e^{i\shalf\epsilon n}\Psi_{n,k-1}\right)
\label{eq:TB}
\end{eqnarray}
where $\epsilon=e_o a b h'/c_o\hbar_{\text{phys}}$ \cite{hb}
is here the dimensionless magnetic flux thru lattice cell. We note that
discrete indices $(n,k)\in\Z^2$ now label the position lattice $(na,kb)$
while continuous indices $(y,x)\in\T^2$ are the conjugate quasi-momenta.
OBI Hamiltonians $\H^\pm$ can be written in terms of $\H$ and its
{\em time-reversal} $\H^* = \H|_{h'\rightarrow -h'}$, namely
$\H^\pm = \H \mp \H^*$, and hence $[\H,\H^*]=0$. Using a different,
Landau gauge $\vec{A}=h'(0,x,0)$ the TB problem (\ref{eq:TB})
can be re-written in terms of 1d Harper equation\cite{Harper}
\begin{equation}
\half\alpha (u_{n+1}+u_{n-1}) +
\beta \cos(n\epsilon-\vartheta)u_n = E u_n.
\label{eq:Harper}
\end{equation}
Let us assume for the moment that $\alpha < \beta$. Then $u_n(\vartheta;E)=u_n$
is a unique exponentially localized eigenfunction (EF) of eq. (\ref{eq:Harper})
which has a dense pure point spectrum, and $\Psi_{n,k}(\vartheta;E) = 
\exp\left(i(\vartheta-\half\epsilon n)k\right) u_n(\vartheta;E)$ is a {\em
degenerate dense} set of EFs of TB problem (\ref{eq:TB}), 
$\H \Psi_{n,k}(\vartheta;E) = E\Psi_{n,k}(\vartheta;E)$,
for a {\em dense} set of parameters $\vartheta$ \cite{Sokoloff}. 
Though $\H$ and $\H^*$ should have a common set of EFs, 
$\Psi_{n,k}(\vartheta;E)$ 
is not an EF of $\H^*$, neither it is in $L^2$ 
since it is {\em extended} in variable $k$.
We search for such EF with an ansatz 
$\Phi_{n,k}(\vartheta;E,E')=\sum_j v_j \Psi_{n,k}(\vartheta+\epsilon j;E)$ 
and require $\H^* \Phi_{n,k}(\vartheta;E,E')=E'\Phi_{n,k}(\vartheta;E,E')$
yielding the Harper equation (\ref{eq:Harper}) for coefficients 
$v_n=u_n(\vartheta;E')$. Thus we obtain a common set of EFs of 
$\H$ and $\H^*$ in terms of a `convolution' of two 1d Harper functions
\begin{equation}
\!\!\!\!\!\Phi_{n,k}(\vartheta;E,E')\!=\!
\sum_l u_{l+n}(\vartheta;E) u_l(\vartheta;E') 
e^{i(\vartheta - \epsilon l - \shalf\epsilon n)k}
\label{eq:EF}
\end{equation}
which is also a common set of EFs of $\H^\pm$ (\ref{eq:Hex})
\begin{equation}
\H^\pm \Phi_{n,k}(\vartheta;E,E') = (E\mp E')\Phi_{n,k}(\vartheta;E,E').
\end{equation}
The property $\Phi_{n,k}(\vartheta+\epsilon;E,E')=\Phi_{n,k}(\vartheta;E,E')$ 
suggests independence of EF on parameter $\vartheta$ provided
$\epsilon/2\pi$ is irrational. If $\alpha > \beta$ localized 
EFs can be constructed analoglously by `duality transformation'
$n\leftrightarrow k,y\leftrightarrow x$. Thus, we found that OBI $\H^\pm$
have a dense pure point spectrum for $\alpha\ne\beta$, hence
XXmh is non-mixing, non-ergodic and even {\em completely
integrable}, namely `zero-energy' eigenstates of $\H^\sigma$
are the images of a complete set of {\em conserved charges},
$Q_\sigma(E) = \sum_{n,k} \Phi^*_{n,k}(E,\sigma E) W^\sigma_{n,k}$,
$[H,Q_\sigma(E)]\equiv 0$. However, time autocorrelation function
$(A(t)|A)$ of a certain observable $A$ may still decay to zero
provided the image function $\Psi^A$ is orthogonal
to all (localized) EFs (\ref{eq:EF}) of $\H^\pm$.
Interestingly, this happens with the spin-current $j_s=W^+_{1,0}-W^+_{-1,0}$
if $\alpha < \beta$ (i.e. $J < h$), since EFs (\ref{eq:EF}) have the
following properties
$\Phi_{n,k}(E,E')^* = \Phi_{n,k}(E',E) = \Phi_{-n,-k}(E,E')$ and
$\Phi_{n,0}(E,E')^* = \Phi_{n,0}(E,E')$ (we put $\vartheta:=0$)
implying $\Phi_{n,0}(E,E') = \Phi_{-n,0}(E,E')$. So we have
$\braket{\Psi^{j_s}}{\Phi(E,E')} \equiv 0$. This proves
$(j_s(t\rightarrow\infty)|j_s)\rightarrow 0$ and non-ballistic
spin-transport (vanishing {\em spin stiffness}
$D_s:=\lim_{T\rightarrow\infty}(1/T)\int_0^T dt(j_s(t)|j_s)$)
for $J < h$, while for $J > h$ we find in general ballistic
transport ($D_s > 0$) since no similar symmetry exists for
the other index $k$.

{\em Example II:} {\em kicked} XXmh model (kXXmh)
with time-dependent Hamiltonian
\begin{equation}
H(t) = J U^+_{1,0} + \half h (U^+_{0,1} + U^+_{0,-1})
\sum_m \delta(t-m).
\end{equation}
One-period propagator from just after the kick
\begin{equation}
U=\exp\left(-i\half h (U^+_{0,1} + U^+_{0,-1})\right)
\exp\left(-i J U^+_{1,0}\right)
\end{equation}
is equivalent to Floquet quantum maps of two kicked OBI
\begin{eqnarray}
\U^+ &=& \exp\bigl({\textstyle\frac{i\beta}{\hbar}}\sin\py\sin x\bigr)
\exp\left({\textstyle\frac{-i\alpha}{\hbar}}\sin\px\sin y\right),\label{eq:U1}\\
\U^- &=& \exp\bigl({\textstyle\frac{-i\beta}{\hbar}}\cos\py\cos x\bigr)
\exp\left({\textstyle\frac{-i\alpha}{\hbar}}\cos\px\cos y\right). \nonumber
\end{eqnarray}
In the following we will consider only the map $\U^+$ since the space
${\frak M}_+$ contain physically more interesting observables.
The Floquet evolution $\Psi^A(m) = \U^{+ m}\Psi^A(0)$ yielding
Heisenberg evolution of observables $A(m)\in{\frak M}_+$ is in
the `classical' limit equivalent to a volume-preserving
$(2\times2)$d map on $\T^2\times\R^2$
\begin{eqnarray}
x' &=& x + \alpha\cos p_x\sin y, \quad p'_y = p_y - \alpha\sin p_x\cos y,
\label{eq:clmap}\\
y' &=& y - \beta\cos p'_y\sin x',\quad p'_x = p_x + \beta\sin p'_y\cos x',
\nonumber
\end{eqnarray}
which is non-integrable and (almost) fully chaotic for sufficiently large
kick parameters, $\alpha,\beta\gg 1$.
Interesting question is now if and when the dynamics of kXXmh is quantum
mixing and how it corresponds to dynamics of the `classical map' 
(\ref{eq:clmap}) as $\hbar=\half\epsilon\rightarrow 0$. This problem has 
been approached numerically by iterating the one-body Floquet map $\U^+$ 
on a finite (truncated) momentum space $(n,k)\in\{-N/2\ldots N/2\}^2$. 
The position states are then discretized as $x_j = s j,y_j = s j, s=2\pi/N$. 
The truncated Floquet map $\U^+$ 
can be efficiently implemented by means of Fast Fourier Transformation
(FFT), namely if $F$ is 1d FFT on $N$ sites then $N^2\times N^2$ 
Floquet matrix is de-composed as \goodbreak
$(F^{-1}\otimes 1)(\text{diag\,}C_{n,k})(F\otimes F^{-1})
(\text{diag\,}D_{n,k})(1\otimes F)$, with diagonal matrices 
$C_{n,k}=\exp\left(i(\beta/\hbar)\sin s n \sin \hbar k\right)$ and
$D_{n,k}=\exp\left(-i(\alpha/\hbar)\sin \hbar n\sin s k\right)$,
requiring $\sim 4 N^2 \log_2 N$ computer operations per time step.
In order to avoid recurrences of quantum probability due to
finiteness of momentum space we use an absorbing boundary in momentum 
space, namely after each iteration of the truncated Floquet
map we multiply the wave-function by a box-window,
$\Psi_{n,k}(m)\rightarrow
\theta(N/2-\alpha/\hbar-|n|)\theta(N/2-\beta/\hbar-|k|)\Psi_{n,k}(m)$,
Convergence to true dynamics on a torus has been checked by comparing results
for different truncations, say $N$ and $N/2$ (we went up to $N=2^{14}$).

In Fig.1 we show numerical results for the auto-correlation function
of the spin current $C(m) = (j_s(m)|j_s(0))$ while similar, compatible
results have been obtained for the time-correlations of other observables.
(i) For sufficiently large kick parameters $\alpha,\beta$ the classical map 
(\ref{eq:clmap}) is strongly chaotic and mixing exhibiting normal diffusion
in momentum plane $(p_x,p_y)$. However, kXXmh is not exactly mixing for any 
finite $\hbar$: $|C(m)|$ is rapidly (possibly {\em exponentially}) 
decreasing down to some value $C^* = \overline{|C(m)|}$ where it saturates. 
When we decrease $\hbar$, $C^*$ decreases proportionally, $C^* \propto \hbar$, 
and so in the `quasi-classical'/continuum limit 
$\hbar=\half\epsilon\rightarrow 0$ the {\em point} spectrum of $\U^+$ vanishes
and kXXmh approaches mixing behaviour in accordance with the map 
(\ref{eq:clmap}). (ii) For smaller but still finite values of $\alpha,\beta$ 
the classical map enters into the regime of KAM quasi-integrability with 
invariant tori suppressing the diffusion of momenta $(p_x,p_y)$. 
Correspondingly, kXXmh is non-mixing and $C^* \sim 1$ for any 
value of $\hbar$. In this regime, $C(m)$ is very weakly $\hbar-$dependent.
In both regimes, (i) and (ii), the square widths of the `wavepackets' 
$\bra{\Psi^{j_s}(m)}\hat{p}^2_{x,y}\ket{\Psi^{j_s}(m)}$ have been found to
be {\em uniformly} increasing in time and limited only by the size of the 
truncated momentum space $N$. This rules out the possibility of {\em quantum 
localization} and existence of {\em pure point} spectrum, and indicates
coexistence of {\em point} and {\em continuous} spectrum for any 
finite $\hbar$ (and finite $\alpha,\beta$), a situation similar to 
(possibly related) 1d kicked Harper model \cite{KH}. In the limit 
$\alpha,\beta\rightarrow 0$, the continuous spectral component vanishes and we
recover integrable XXmh model with {\em pure point} spectrum as discussed above.
The quantum correlation function $C(m)$ seem to follow the
quasi-classical propagator only up to logarithmically short time, namely we 
found empirically that deviation (when it is small) increases exponentially 
$|C(m)-C_{\hbar\rightarrow 0}(m)|\approx 0.022 \hbar^2 e^{\lambda m}$ 
with $\lambda\approx 0.59$ for $\alpha=3,\beta=0.75$ and
$\lambda\approx 1.1$ for $\alpha=6,\beta=1.5$.

{\em Conclusions.} 
In a specific $\infty$d class of (Pauli spin, or spinless fermion) quantum
field models in 1d, the Heisenberg time evolution in two disjoint $\infty$d
linear subspaces of essential field observables has been shown to be formally
equivalent to the Schr\" odinger dynamics of a class of one-body image problems 
on a 2d torus (or 2d lattice). Autonomous models of this class were found to be
completely integrable, pointing out a novel class of integrable one-body 
problems (\ref{eq:H1},\ref{eq:com}). For example, dynamics of XX chain in a 
static quasiperiodic transversal field has been solved in terms of Harper 
equation\cite{Satija}. However, time-dependent (e.g. periodically kicked) 
models of our class behave in a non-integrable fashion being mapped onto 
one-body problems with chaotic classical limit. It seems that spatial 
modulation is crucial to break integrability since spin chain kicked with
{\em homogeneous} transversal field remains completely integrable as found in
\cite{KI}. 
In the {\em contunuum field limit} our kicked spin chain model (kXXmh) has been 
demonstrated to undergo a (phase) transition from mixing to non-mixing dynamics
(similar to a transition found in \cite{KtV}), as its one-body counterpart in 
the {\em classical limit} undergoes a stochastic transition from chaotic to 
quasi-regular motion. This is an interesting link between quantum field theory 
and chaotic dynamics and should inspire
future research in this direction.
Such approach to long-time dynamics of certain (non-integrable) quantum
many-body systems, since it makes time evolution formally equivalent to
`quantum chaos' in few degrees of freedom, overcomes the traditional problems
due to huge Fock space in thermodynamic limit.
Financial support by the Ministry of Science and Technology of R. Slovenia is
acknowledged.

\begin{figure}[htbp]
\begin{center}
{\leavevmode
\epsfxsize=3.33truein
\epsfbox{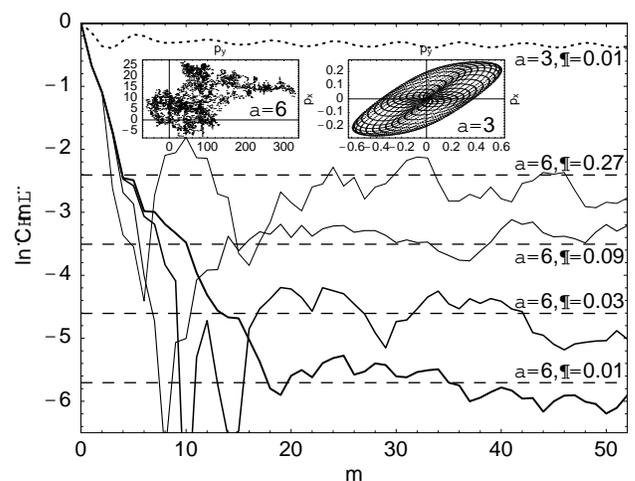}}
\end{center}
\vspace{-2mm}
\caption{$\ln|C(m)|$ in kXXmh for several indicated values of
parameters $\alpha$ and $\epsilon$ while $\beta = \alpha/4$.
Note the transition to mixing dynamics as $\epsilon$ is decreased 
(heavy--light full curves) for `chaotic case' $\alpha=6$, and stable
non-mixing behaviour for `quasi-regular' case $\alpha=3$ (dotted curve,
curves for other (small) values of $\epsilon$ are almost indistinguishable). 
Broken lines at $\ln|\epsilon/3|$ indicate the scaling $C^* \propto \hbar = 
\half\epsilon$. In two insets we show two orbits in momentum plane (chaotic, 
diffusive for $\alpha=6$ and quasi-regular for $\alpha=3$) of the map 
(\ref{eq:clmap}) of length $3000$ starting at 
$x_0=0.2,y_0=0,p_{x0}=0,p_{y0}=0.005$.}
\label{fig:1}
\end{figure}


\begin{references}
\bibitem{hb} $\hbar$ is a dimensionless model parameter which has to be
distinguished from the physical Planck constant $\hbar_{\text{phys}}$.
\end{references}
\end{document}